\documentclass[
  aps,
  prl,
  reprint,
  amsmath,
  amssymb,
  superscriptaddress,
  nofootinbib,
  longbibliography,
  floatfix
]{revtex4-2}

\usepackage{graphicx}
\usepackage{bm}
\usepackage{hyperref}
\hypersetup{
  hidelinks,
  pdftitle={Symmetry-agnostic stellarators for collisionless confinement},
  pdfauthor={W. Sengupta, A. Bhattacharjee, S. Buller}
}

\newcommand{\GammaW}{\Gamma_{\mathsf W}}
\newcommand{\dd}{\mathrm{d}}
\newcommand{\B}{\mathbf{B}}
\newcommand{\cE}{\mathcal E}
\newcommand{\cJ}{\mathcal J}
\newcommand{\cB}{\mathcal B}
\newcommand{\Br}{B_{\rm r}}
\newcommand{\dpsi}{\partial_\psi}
\newcommand{\dal}{\partial_\alpha}
\newcommand{\dl}{\Delta\ell}
\newcommand{\vpar}{v_\parallel}

\begin{document}

\title{Symmetry-agnostic stellarators for collisionless confinement}

\author{W. Sengupta}
\email{wsengupta@princeton.edu}
\affiliation{Department of Astrophysical Sciences, Princeton University,
Princeton, New Jersey 08544, USA}

\author{A. Bhattacharjee}
\email{amitava@princeton.edu}
\affiliation{Department of Astrophysical Sciences, Princeton University,
Princeton, New Jersey 08544, USA}

\author{S. Buller}
\email{sb0095@princeton.edu}
\affiliation{Department of Astrophysical Sciences, Princeton University,
Princeton, New Jersey 08544, USA}

\begin{abstract}
Quasisymmetry, omnigenity and piecewise omnigenity confine trapped particles by
making the bounce action independent of the field-line label. Recent
optimizations produce mixed-symmetry stellarators that confine alpha particles
well without them. We propose a general theory for them. From Whitham modulation
theory we define iso-action, which requires only that the drift surface close
and allows misalignment with flux surfaces. A solvable model supplies an exact
relation between trapped segments while branch actions vary. We develop a proxy
$\GammaW$ for the reach that misalignment costs.
\end{abstract}

\maketitle

Stellarators confine charged particles in a fully three-dimensional magnetic field without the large externally driven plasma currents needed for a tokamak. That freedom comes at a price: the loss of axisymmetry as well as canonical angular momentum conservation makes single-particle confinement more challenging. In a generic stellarator, trapped particles generally drift radially over many bounce periods. Fusion-born alpha particles can thus be lost before their energy heats the plasma. To prevent such loss, additional geometric constraints must be imposed on the magnetic field.

The collisionless bounce-averaged drifts follow from the second adiabatic invariant
\begin{equation}
 J(\cE,\mu,\psi,\alpha)=\oint \vpar\,\dd\ell,
 \label{eq:Jdef}
\end{equation}
where $\cE$ is the energy per unit mass, $\mu$ the magnetic moment per unit mass, $\psi$ labels the flux surface, and $\alpha$ labels the field line on that surface. 
A field is said to be omnigenous (O)\cite{cary1997,Hall1975,helander2014} when $J$ is the same on every field line,
\begin{equation}
(\dal J)|_{\cE,\mu,\psi}=0,
\label{eq:omn}
\end{equation}
which ensures that the bounce-averaged radial drift vanishes. There is a ladder of ways to
meet this condition, each a relaxation of the one before. Quasisymmetry (QS)~\cite{nuhrenberg1988} 
is the simplest: the field strength looks the same along every
line (up to a phase shift), so $J$ cannot depend on the line. Omnigenity (O) relaxes this. The
wells may differ from line to line, provided each keeps its action. Piecewise
omnigenity (pwO)~\cite{velasco2024} relaxes it further, letting a particle change
trapped class, provided the action is fixed within each class.

Every rung of this ladder holds the action of each occupied well fixed. We show in this Letter 
that this restriction is not necessary, giving us greater flexibility in design. Suppose that, as it drifts, a
particle visits several wells in turn. Each leg of the journey may carry a finite
radial drift, and each well's action may vary from line to line, yet the drift
accumulated over the \emph{complete sequence of wells} can still cancel. We call
the motion \emph{iso-action} when the orbit-selected action, fixed by that
sequence and not chosen independently, is single-valued after one turn and its
action contour closes inside the plasma. What matters for practical
confinement is the fraction of trapped phase space whose drift surfaces close
inside the plasma, and design should maximize that fraction.
Iso-action is strictly weaker than pwO. Indeed, QS, O, and
pwO are all special cases, obtained when the individual actions are separately
fixed. The iso-action principle, which builds on a key insight of Whitham's for slowly-varying continuous media \cite{whitham1974}
and Dewar's seminal application of it to magnetohydrodynamics (MHD) \cite{dewar1970},
provides the foundation for symmetry-agnostic stellarators proposed in this Letter.
Whitham's insight is that the adiabatic invariant of a fast--slow oscillation
system is conserved along the characteristics of the modulation, not
pointwise. Thus, the trapped-particle bounce action need only be constant
along the drift, not on every field line.

Recent optimizations provide the practical motivation for and demonstration of the iso-action principle.
Direct fast-ion optimization can produce fields with good energetic-particle
confinement without exact QS or O or pwO
\cite{bindellandremanpadidar2023,paul2022}. Flat-mirror optimization gives
another route to reduced radial drift \cite{velasco2023flatmirror}. These studies demonstrate that local action constancy 
is \emph{not} necessarily the most useful design target.

The new constructions impose orbit closure at different levels of the ladder we describe above. For QS and
O, banana tips move on equal-$|\B|$ contour branches that wind
around the surface. The projected guiding-center (g.c.) orbit is carried by those
$B=|\B|$ contours, but it is not generally one constant-$B$ contour.
pwO separates g.c. orbit closure from $B$
contour closure. We take the next natural step and retain only closure of the
drift loop, defined below.

\textit{Bounce motion and the reduced action.} A trapped particle bounces rapidly along a field line and drifts slowly across
field lines. At fixed energy and magnetic moment, the fast motion is measured
by the reduced one-transit action
\begin{equation}
 \cJ(\psi,\alpha;\Br)=\int_{\ell_-}^{\ell_+}
 \sqrt{\Br-B(\psi,\alpha,\ell)}\,\dd\ell.
 \label{eq:JdefBr}
\end{equation}
Here $\psi$ labels a flux surface, $\alpha$ labels a field line, and
$B=\Br$ at the bounce points $\ell_\pm$, with $\Br=\cE/\mu$ and
$J=2\sqrt{2\mu}\,\cJ$. Equation~\eqref{eq:JdefBr} can be thought of as an optical path along
the trapped segment, with effective refractive index $n_{\rm eff}\propto\sqrt{\Br-B}$
vanishing at the bounce points. Read that way, $\dal\cJ\,d\alpha$ is the
transverse optical-path difference between neighboring field lines. The
bounce time is proportional to
$\partial_{\Br}\cJ$. The bounce-averaged radial drift is proportional to
$-\dal\cJ/\partial_{\Br}\cJ$. The drift in $\alpha$ is proportional to
$\dpsi\cJ/\partial_{\Br}\cJ$. Their ratio is independent of the bounce-time
normalization. Bounce-averaging therefore transports the action along the slow
motion. This is the Whitham reduction of the bounce--drift problem
\cite{whitham1974}, where the bounce is the fast cycle and the field-line label is
the slow variable.
The physical fast cycle selects the action, and its total derivative vanishes:
\begin{equation}
 0=\frac{\dd\cJ_\Gamma}{\dd\alpha}
 =\dal\cJ_\Gamma+\frac{\dd\psi}{\dd\alpha}\dpsi\cJ_\Gamma,
 \qquad
 \frac{\dd\psi}{\dd\alpha}=-\frac{\dal\cJ_\Gamma}{\dpsi\cJ_\Gamma}.
 \label{eq:transport}
\end{equation}
The solution curve in the $(\psi,\alpha)$ plane is the Whitham
characteristic. The transported action is constant along it, so it is equally
a contour of that action, and we call it an action contour. The
subscript $\Gamma$ records the wells, their order, and
their signed traversal numbers as selected by the physical orbit. Unlike pwO, which fixes each branch action, iso-action constrains only the combination the orbit selects. A well can split, merge, or disappear during one
turn in $\alpha$. The action is transported along the slow drift through
the total-derivative law $\dd\cJ_\Gamma/\dd\alpha=0$, which does not require
$\dal\cJ_\Gamma=0$. 

We call the path that the bounce center follows over one turn in $\alpha$,
through the wells the orbit selects, the \emph{drift loop}. It is iso-action when
the transported action is single-valued after that turn and the corresponding
action contour closes inside the plasma. The action contour
may be corrugated relative to the flux surfaces. Closure then bounds the radial
motion instead of eliminating it locally. A drift loop that reaches the last
closed flux surface is lost. One that closes inside it has finite radial
reach, by which we mean the largest radial excursion it makes from its
starting surface.

For a single well that persists over one turn in $\alpha$, closure can be
tested directly from the field. Let
$\dl(\cB)=\ell_+(\cB)-\ell_-(\cB)$ be the distance between the two points with
$B=\cB$. Integrating by parts, with $B_{\rm floor}$ any fixed constant below
the well bottom, the action is
$\cJ=\tfrac12\int_{B_{\rm floor}}^{\Br}\dl(\cB)\,\dd\cB/\sqrt{\Br-\cB}$, a
functional of the chord widths alone \cite{senguptaweitzner2018}. Fixing every
chord width in $\alpha$ is the Cary--Shasharina route to O
\cite{cary1997}, the limit in which the transverse path difference vanishes.
When a single well persists through the turn and the same pitches are
available throughout, Abel inversion makes this equivalent to $\dal\cJ=0$.
Differentiating the same Abel representation in radius gives
\begin{align}
 \dpsi\cJ
 &=-\frac12\int_{B_{\rm floor}}^{\Br}
 \frac{Q(\cB)}{\sqrt{\Br-\cB}}\,\dd\cB,
 \qquad Q=-\dpsi\dl, \notag\\
 Q(\cB)
 &=\sum_{\pm}\frac{\dpsi B(\ell_\pm)}
 {|\partial_\ell B(\ell_\pm)|}.
 \label{eq:paired}
\end{align}
The kernel is positive. If $Q$ does not change sign, $\dpsi\cJ$ does not either at every trapped pitch in that well, and the particles cannot reverse their precession
\cite{rodriguez2024maxJ}. This is weaker than demanding a favorable
radial gradient at every point because one flank of the well may compensate
the other after the geometric weighting in Eq.~\eqref{eq:paired}. However, the sign
condition alone does not guarantee action confinement.

\textit{Auxiliary Schr\"odinger problem and the role of finite-gap potentials.}
The bounce-adiabatic ordering makes the action invariant only to exponential
accuracy. On an infinite line there are reflectionless wells for which the
change in $\cJ$ vanishes identically instead \cite{GjajaBhattacharjee1992}. A
field line in a stellarator does not provide that limit, because the potential $B$ is periodic on a rational surface and quasiperiodic on an irrational
one. The periodic problem is reached by attaching to the parallel motion the
auxiliary Schr\"odinger equation
\begin{equation}
 \left[-\varepsilon^2\partial_x^2+U\right]y=\Lambda y,
 \qquad U=B_M-B,
 \label{eq:hill}
\end{equation}
with $B_M$ the largest field strength on the surface and $\varepsilon$ a WKB
smallness parameter that mimics the fast variation along the field line.
Writing $y\sim\exp(iS/\varepsilon)$ gives $(\partial_xS)^2=\Lambda-U$, and the
choice $\Lambda=B_M-\Br$ gives $\Lambda-U=B-\Br$. The semiclassical limit is
therefore the g.c. picture, and the two problems share their turning points
and critical levels. They run over complementary real cycles, the auxiliary
phase being real where $B>\Br$ and the bounce integral where $B<\Br$. A generic $U$ opens infinitely many
spectral gaps, whereas a finite-gap $U$ opens finitely many, and the finite-gap
potentials are the periodic analogues of the reflectionless ones
\cite{Novikov1984,gesztesyweikard1995}. However, we note that a
finite-gap potential is not reflectionless, so it does not inherit the exact
cancellation of the infinite-line construction, which belongs to the limit
$\ell\to\infty$. Finite-gap potentials are useful because they can approximate
any smooth periodic potential \cite{marchenkoostrovskii1975}. They furnish a
handful of parameters that fix the well shape, the width at every field
strength, and the bounce action across a whole band of pitches. Each interior
maximum of $B$ marks one of a finite number of critical field strengths at
which orbit classes merge or separate.

\textit{A solvable multibranch model.} Inspired by finite-gap potentials,
we now consider an analytically tractable model in which the orbit changes
well during a turn. Brizard
and Westland showed that an asymmetric double well can hold a particle
for the same time in either lobe at a given energy~\cite{brizardwestland2017}.
The Treibich--Verdier B3 potential~\cite{treibichverdier1992,gesztesyweikard1995} carries that
property into a model field-strength profile along a field line,

\begin{subequations}
\label{eq:b3}
\begin{align}
B_{\rm B3} = 6k^{2}\,\mathrm{sn}^{2}(x\mid k^{2})
   + 2k^{2}\,\mathrm{sn}^{2}(x+K+\chi\mid k^{2}),
\label{eq:b3pot}\\
\mathcal{J}_{D}-\mathcal{J}_{S} = \pi(\sqrt{6}-\sqrt{2})\equiv\mathcal{J}_{*},
\label{eq:b3id}
\end{align}
\end{subequations}
with $k^{2}=0.8$ and period $2K$ in both $x$ and the model field-line
label $\chi$, where $\alpha=\pi/3+\pi\chi/(3K)$. As $\chi$ advances, a
well splits into a deep and a shallow daughter and merges again
\cite{caryescandetennyson1986,neishtadt1986}.
Both daughter actions vary smoothly with $\chi$, so the field is not
pwO. The two wells have equal bounce times at every energy $u$,
$\partial_{u}\mathcal{J}_{D}=\partial_{u}\mathcal{J}_{S}$,
so $\mathcal{J}_{*}$ is independent of energy. It is independent of
$\chi$ as well. The consequence for the drift loop is that whichever daughter well the orbit
goes through at a split generates the same radial characteristic, and the
drift cancels once the g.c. has completed the full merge, split and remerge
sequence.

\begin{figure*}[t]
\centering
\includegraphics[width=0.32\textwidth]{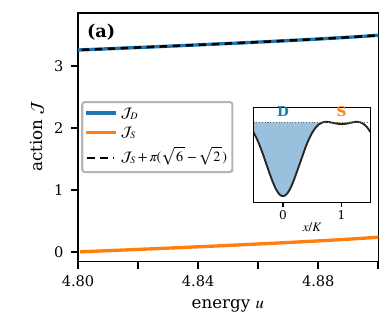}\hfill
\includegraphics[width=0.32\textwidth]{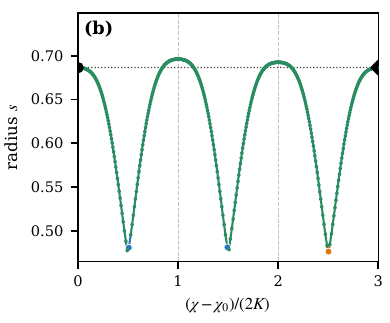}\hfill
\includegraphics[width=0.32\textwidth]{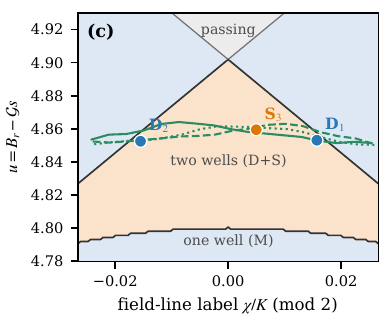}
\caption{\label{fig:b3}%
\textbf{The B3 drift loop closes without a constant-$|\B|$ contour.}
(a) Both branch actions of Eq.~\eqref{eq:b3pot} vary with energy at the same
rate, so their difference is pinned at the constant $\cJ_*$ of
Eq.~\eqref{eq:b3id}. The pinning also holds as the field line varies. Neither
action is then separately conserved, so the field is not pwO. Inset: the deep (D) and shallow (S) trapped segments at
$\chi=0$. (b) Direct fast--slow integration confirms the closure: the bounce
center returns to its starting section after three B3 periods, its
section displacement staying below $2\times10^{-6}$. This is numerical evidence in the
B3 torus embedding, not a guiding-center orbit in a realized three-dimensional
MHD equilibrium. (c) Well topology near the split window, in the
plane of field-line label and $u=\Br-\mathcal{G}s$, with $\mathcal{G}$ the mean
radial gradient of the field strength: one well
(M), two wells (D+S), passing. The drift loop of (b) (three passes: solid,
dashed, dotted) enters the split region once per period and takes the deep or
the shallow daughter, D$_1$, D$_2$, S$_3$.}
\end{figure*}

The above is verified in Fig.~\ref{fig:b3}. Panel (a) evaluates the two branch
actions directly from Eq.~\eqref{eq:b3pot} using numerical integration. Each $\cJ$ varies with energy, at the same rate, and their difference is $\pi(\sqrt6-\sqrt2)$ as Eq.~\eqref{eq:b3id} requires.
Panel (b) shows the trajectory of the g.c. particle in $s$ and $\chi$ over a loop. 

Panel (c) shows how the orbit from (b) transitions between regions with single or double wells.
Once per period the loop enters the region where the well is split and bounces in the deep or the
shallow daughter well. This particular orbit traverses the wells in order D, D and S, before the daughter wells
remerge. The points of transition are also marked in (b) with the same colors.
The action of the occupied well is far from constant along a curve of
fixed field strength, varying by $40\%$ of its launch value at $u=4.0$
across one period below the split band, so the field is not
piecewise omnigenous and the drift cancels only over the full orbit. Closure is
therefore a property of the action contour and not of any contour of $B$, which
is the iso-action condition.

\textit{Single-well closure in optimized fields.}
The above analytical example relied on multiple wells producing the same radial drift. For the case of a single well, the iso-action concept can be evaluated directly from the magnetic field without orbit-tracing, because the well never splits and the question of which daughter the orbit enters never arises. The drift-loop can simply be evaluated from contours of $\cJ$ within the single well~\cite{goodman2024prxe}.

We calculated $\cJ$ for the five alpha-particle-optimized configurations
of Ref.~\cite{landreman2026bayesian}. In all five,
$\partial_{\alpha}\mathcal{J}$ is smooth and nonzero over the radii
examined, with peak-to-peak action modulation across field lines ranging
from $0.9\%$ to $56\%$. By definition, therefore, none of them
is omnigenous or pwO.

Following one occupied well in each, we constructed the action contour
$\mathcal{J}(s,\alpha)=\mathcal{J}_{0}$ from
the magnetic field alone, with no orbit integration. Every contour
closes inside the plasma, and the independently traced g.c. orbit lies
on it. In the strongest modulation case the orbit-contour residual is
$0.38\%$ of the orbit width, while the action is conserved along the
orbit to about $1\%$.
The five split into helical and mirror
well trains, and $\partial_{\psi}\mathcal{J}$ reverses sign between the
two groups, so a fixed maximum-$J$ sign is not a universal confinement
label. What the groups do share is an action contour that closes inside the 
plasma, which is the iso-action condition. These
are single-branch tests, and they do not probe the multibranch
cancellation of Eq.~(\ref{eq:b3id}).

\textit{Radial reach and the proxy $\GammaW$.}
As discussed above, closed drift 
surfaces need not coincide with flux surfaces. Their misalignment produces a finite radial excursion. For a marker launched at
normalized flux-surface label $s_{0,i}$, let $s^{\mathsf W}_{\max,i}$ be the largest radius reached
along the action contour of Eq.~\eqref{eq:transport} over a specified time. We define the
Whitham energetic-particle proxy
\begin{align}
 R_{{\mathsf W},i}
 &=\min\!\left\{1,\max\!\left[0,
 \frac{s^{\mathsf W}_{\max,i}-s_{0,i}}{1-s_{0,i}}\right]\right\}, \notag\\
 \GammaW&=\frac{\sum_i h_iR_{{\mathsf W},i}}{\sum_i h_i}.
 \label{eq:GammaW}
\end{align}
where $h_i$ is the alpha birth weight. It carries both the integration time
and the remaining distance to the plasma boundary. The excursion is evaluated
along the bounce-averaged characteristic of Eq.~\eqref{eq:transport}. It is
computed from the field alone and not by tracing guiding-center orbits. The
occupied well is followed by continuity through its splits and merges. The
branch taken at a split is therefore assigned by a deterministic rule and not
by the capture dynamics. $\GammaW$ does not include beyond-g.c. physics such as
finite-orbit-width effects.

The Nemov metrics $\Gamma_w$ and $\Gamma_c$ \cite{nemov2005,nemov2008} are
energetic-particle metrics built from the same derivatives of $\cJ$, evaluated
on a single flux surface. $\GammaW$ instead integrates the action contour across
surfaces, converting misalignment into a radial reach.

\begin{figure*}[t]
\centering
\includegraphics[width=0.95\textwidth]{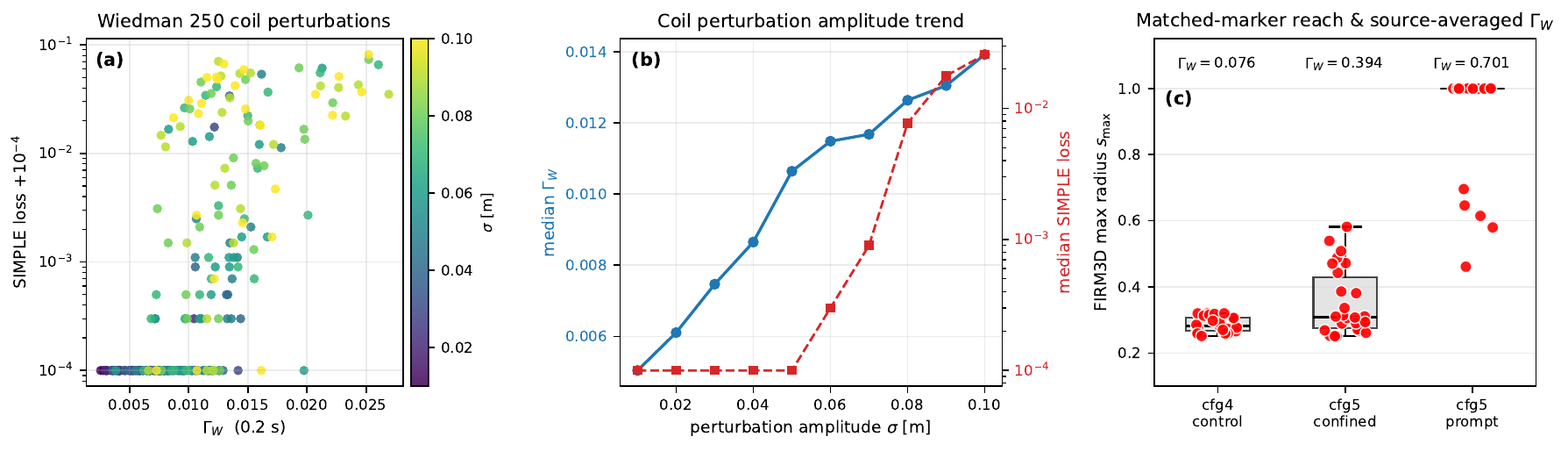}
\caption{\label{fig:gammaw}%
\textbf{Tests of $\GammaW$.} (a) Across 250 coil perturbations $\GammaW$
correlates with SIMPLE loss, while the QS error remains stronger. (b) The
bin-median $\GammaW$ follows perturbation amplitude. (c) FIRM3D \cite{paul2026firm3d} maximum radius for three matched
cohorts of twenty-six markers in cfg4 and cfg5, labeled by the
source-averaged $\GammaW$ of each cohort.}
\end{figure*}

Figure~\ref{fig:gammaw} tests the metric at three levels. In 250 coil
perturbations of a precise QH configuration \cite{landremanpaul2022,wiedman2024coil},
$\GammaW$ has Spearman correlation $0.656$ with SIMPLE losses
\cite{ALBERT2020109065}. The QS error reaches $0.736$.
In panel (b), the bin-median
$\GammaW$ rises monotonically with perturbation amplitude, so much of the
pooled trend is amplitude-driven. Across amplitude bins it tracks the
bin-median loss at $0.985$.

Figure~\ref{fig:gammaw}(c) compares individual traced trajectories rather than whole
configurations. The promptly lost,
confined, and near-omnigenous cohorts of 26 matched markers have mean
reaches $0.701$, $0.394$, and $0.076$. Direct trajectories confirm that the
prompt markers remain in a single well. These results support $\GammaW$ as an
excursion estimator in the tested adiabatic regime. On the five
alpha-optimized configurations of Ref.~\cite{landreman2026bayesian}, the
action contour reproduces the traced radial reach within $1\%$.

In the Paul reactor-scale set \cite{paul2022},
$\GammaW$ gives its largest value to the lossy quasihelical case but
under-ranks the lossy quasiaxisymmetric ones, whose trapped-banana loss needs
finite-orbit-width information. That quasihelical case itself lies outside the
adiabatic regime over the quoted window.

\textit{Discussion.}
Stellarator confinement can be achieved through closing the drift-surface, rather than eliminating the flux-surface-averaged radial drift.
The B3 model shows that this closure can
survive when the branch actions vary and the equal-$|\B|$ contours form
closed loops rather than winding around the surface.
$\GammaW$ then measures the radial reach of the residual misalignment. It is
obtained from the action contour and not from full guiding-center orbit
tracing.

The result suggests a symmetry-agnostic optimization hierarchy of two
requirements. The orbit-selected characteristic must return to its starting
branch and section. Its radial reach must not exit the confined region. QS
and O satisfy both by construction, which is why they have never had to be
distinguished. Relaxing local action constancy separates them, and each
must then be imposed on its own.

We remark that while iso-action, symmetry-agnostic stellarator design is a promising path forward, it 
needs to be validated by further simulations and experiments. Exact omnigenity retains the stronger 
property of removing the stellarator-specific low-collisionality $1/\nu$ transport~\cite{helander2014}, whereas
iso-action targets collisionless closure only. At small but finite
collisionality the same misalignment it tolerates produces a small
nonzero neoclassical flux. Thermal and energetic-particle measures are
therefore not interchangeable. Across the five configurations, effective ripple at the launch surface spans
$0.4\%$ to $1.5\%$. The lowest-ripple case has the largest alpha loss. The depth-specific drift cancellation found by Mynick \textit{et
al.}~\cite{mynick1982} earlier is a precedent for that separation.

The bounce-averaged reduction used throughout has a limited domain of
validity. It does not cover the trapped-passing layer,
bounce-precession resonances, drift islands near rational surfaces, or
finite-orbit-width loss, and a separatrix crossing changes the adiabatic
invariant \cite{caryescandetennyson1986,neishtadt1986}. Adiabaticity, assumed
over the longer time scale of multiple bounces, can fail.

The exact B3 identity of Eq.~(\ref{eq:b3id}) holds on the trapped segments. Within that domain, iso-action identifies closure of the
orbit-selected drift loop, rather than local symmetry, as the invariant
design target.

\begin{acknowledgments}
The authors thank E. J. Paul, E. Rodriguez, and M. Landreman for helpful
comments and suggestions. This research is supported by the Department of Energy Award No. DE-SC0024548.
\end{acknowledgments}

\bibliography{references}

\end{document}